\documentclass[aps,prl,reprint,superscriptaddress]{revtex4-2}

\usepackage{dcolumn}
\usepackage{graphicx}
\usepackage{subfigure}
\usepackage{textcomp}
\graphicspath{{figures/}}

\begin{document}
	
	
	\title{Longitudinal Compression of Macro Relativistic Electron Beam}
	
	
	\author{An Li}
	\affiliation{Department of Engineering Physics, Tsinghua University, Beijing 100084, People’s Republic of China}
	
	\author{Jiaru Shi}
	\affiliation{Department of Engineering Physics, Tsinghua University, Beijing 100084, People’s Republic of China}

	\author{Hao Zha}
	\email{ZhaH@mail.tsinghua.edu.cn}
	\affiliation{Department of Engineering Physics, Tsinghua University, Beijing 100084, People’s Republic of China}
	
	\author{Qiang Gao}
	\affiliation{Department of Engineering Physics, Tsinghua University, Beijing 100084, People’s Republic of China}
	
	\author{Liuyuan Zhou}
	\affiliation{Department of Engineering Physics, Tsinghua University, Beijing 100084, People’s Republic of China}
	
	\author{Huaibi Chen}
	\affiliation{Department of Engineering Physics, Tsinghua University, Beijing 100084, People’s Republic of China}
	
	
	\date{\today}
	
	\begin{abstract}
		
		We presented a novel concept of longitudinal bunch train compression capable of manipulating relativistic electron beam in range of hundreds of meters. This concept has the potential to compress the electron beam with a high ratio and raise its power to an ultrahigh level. The method utilizes the spiral motion of electrons in a uniform magnetic field to fold hundreds-of-meters-long trajectories into a compact set-up. The interval between bunches can be adjusted by modulating their sprial movement. The method is explored both analytically and numerically. Compared to set-up of similar size, such as chicane, this method can compress bunches at distinct larger scales and higher intensities, opening up new possibilities for generating beam with ultra-large energy storage.

	\end{abstract}
	
	
	\maketitle
	

	Ultrahigh pulsed power supply has remained a bottleneck in numerous frontier studies in recent years, including fuel pellet compression and ignition in fusion research\cite{FastIgnition,ZPinch,ECRH}, high power wake-field stimulation for novel accelerators\cite{WakefieldAcc,WakefieldAcc2,WakefieldAcc3}, and ultrahigh dose rate X-Ray generation for FLASH radiotherapy or FLASH radiography applications \cite{FLASHRadiotherapy0,FLASHRadiotherapy,DAHRT,FLASHRadiography}.

	To generate pulsed power, energy could be stored in mediums such as laser\cite{CPA,LaserCompression}, electric current\cite{SolidModulator,SolidModulator2}, microwave\cite{MicrowaveModulator,MicrowaveModulator2}, or electron beam\cite{MicrowaveModulator,MicrowaveModulator2}, then get released in short pulse duration. Among these options, electron beam has certain advantages, such as high energy conversion efficiency from AC power, and negligible energy dissipation during compression. Furthermore, as space chage effect decreases with increased particle energy $E$, the power capacity of electron beam is proportional to $E^3$. This indicates an ultrahigh energy storage limit for relativistic beam.
	
	Nevertheless, directly accelerating an high current electron beam to the required energy is both technically challenging and expensive. Induction accelerators used in scientific installations for Z-Pinch or FLASH radiography are typically hundreds of meters long\cite{CLIC2,DAHRT,ZPinch2}, and require a large number of klystrons for power supply. It's more economical to accelerate a long pulse beam with an applicable current, and then compress it to a short duration for power multiplication.

	Electron beam compression schemes that operate at femtosecond or picosecond timescales have been extensively investigated lately\cite{VelocityBunching,MagnetCompress,LaserCompress,CoulombCompress}. While compression for macro beam in nanoseconds to microseconds duration is less mentioned. For relativistic particle, the trajectory span hundreds of meters in spatial scale, resulting in high costs and large set-up dimensions. For example, the Compact Linear Collider (CLIC) study\cite{CLIC2} proposed a bunch train combination system with combiner ring circumference of up to 438 meters.
	
	\begin{figure}[htbp]
		\includegraphics[width=8.5cm]{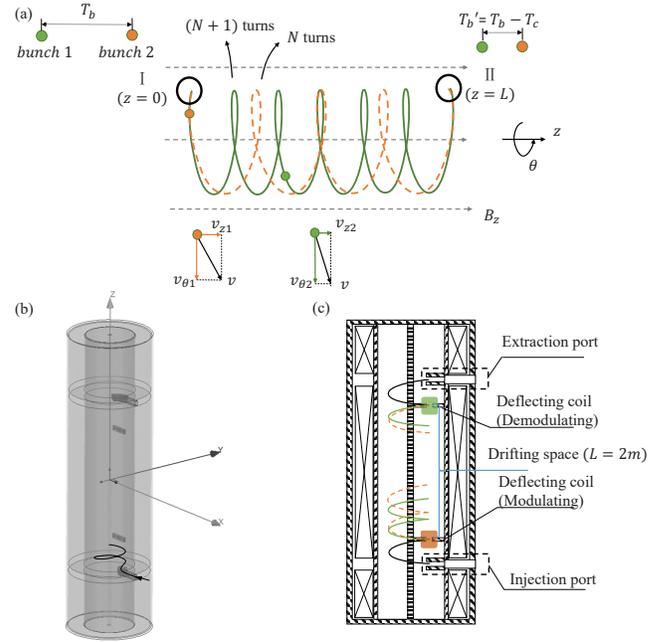}
		\caption{ (a) Sketch of the compression concept. Bunch 1 and 2 spiral with axial velocity components denoted by $v_{z1}$ and $v_{z2}$ respectively. The time interval between bunches at position \uppercase\expandafter{\romannumeral1} is $T_b$, and becomes $T_b' = T_b - T_c$ at position \uppercase\expandafter{\romannumeral2}, where $T_c$ represents the turning period of the bunches. (b) Compression set-up model. The set-up comprises an exterior solenoid, injection and extraction ports, and several deflect magnets inside the cavity. (c) profile of the compression set-up in front view. The modulating and demodulating magnet adjust the axial velocity of the bunches by bending their orbits.}
		\label{FIG.1}
	\end{figure}
	
	In this letter, we present a novel method of compressing a relativistic electron beam in a nanosecond to microsecond timescale, with installation in a spatial scale of several meters.

	The concept is to delay particle bunches by inducing spiral movements in an uniform magnetic field.  As depicted in FIG.\ref{FIG.1}(a), an upstream bunch in the train has a smaller pitch and takes a longer time to travel along the direction of field. A downstream bunch with larger pitch is less delayed and catches up. A beam with a duration of 1 us corresponds to a trajectory length of 300 meters. The trajectory can be spirally folded into a compact cylindrical cavity with a diameter of 1 meter and a length of several meters.

	Regarding the set-up, we use a solenoid to generate an axial-oriented, static, and uniform field $B_{z0}$ within its interior cavity. We employ a combination of permanent magnets as injection and extraction ports, matching injected bunches with the compression set-up.	As shown in FIG.\ref{FIG.1}(b), the beam is externally injected into the cavity with an axial component of velocity $v_{z0}$. The modulating magnet coil generates a time-varying local radial field $B_r$, which provides a delay-dependent deflect effect $\Delta v_z$ and adjusts the spiral pitches of the bunches. The modulated bunch train continues to spiral in the drift space. Time structure of the bunch train is rebuilt at the end of it. The demodulating coil sorts the compressed beam and prepares it for extraction.
	
	
	We illustrate three different compression schemes of the concept, namely spiral pitch modulation, bunch confinement and energy modulation. The compression capability is verified with particle dynamics simulation. Compression of electron bunches with average energy of $6.5\; MeV$ and charge of $1 \;  nC$ in 3 patterns are demonstrated. A compressed beam with a current of $1.8 \; A$ and energy of $65 \;  mJ$ in nanoseconds duration is formed for principle verification.

	\section{Spiral pitch modulation}
	After injection, the bunches get deflected by the modulating magnet. The spiral pitch increases from the leading bunch to the trailing bunch. The pitch of i-th bunch should fit the form of $l_i=\frac{L}{N_i}$, where $N_i$ is the number of spiral turns of the i-th bunch in the drifting space. The bunches then gather again at the demodulating magnet, with a shortened interval of $T_b'=\vert T_b-\Delta NT_c\vert$. The time varying-strength of the demodulating magnet recovers the bunches' axial velocity before extraction.

\begin{figure}[htbp]
		\includegraphics[width=8.4cm]{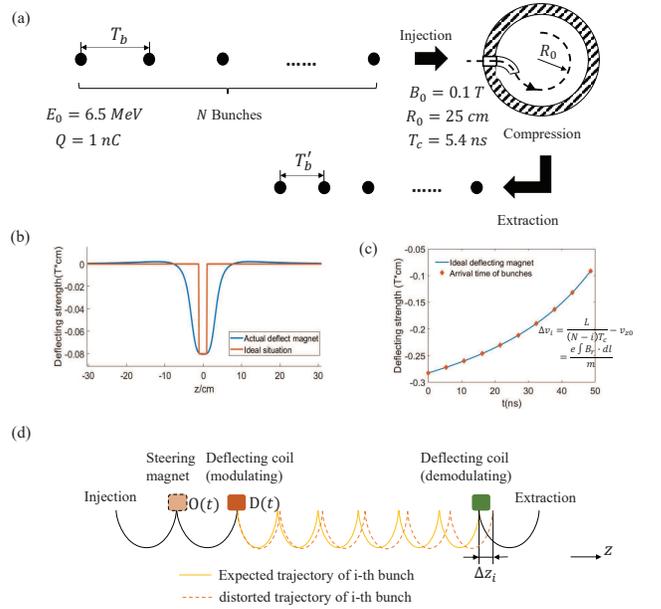}
		\caption{ (a) Compression parameters of the bunch train and the uniform field (b)Deflecting strength distribution of actual and ideal deflect magnet along the z-axis. (c)Time-varying deflecting strength in ideal situation. (d) Sketch of compression system and trajectory of i-th bunch. The solid curve refers to the ideal trajectory and the dased curve refers to the fringe field affected trajectory.}
		\label{FIG.2}
	\end{figure}

	We conducted beam dynamic simulations to investigate the compression concept. Main parameters of the compression are illustrated in FIG.\ref{FIG.2}(a). $T_b=6 \; ns$ is chosen as initial bunch interval. The cavity is 3 meters long with a diameter of $80 \; cm$, and the drifting space occupies 2 meters in length. Axial velocity variation of the i-th bunch is given as $\Delta v_i=\frac{L}{((N-i)T_c)}-v_z0$. With $L=2 \; m$,$N=17$  and  $v_z0=5.8\times10^7 \; m/s$,  a variation of $\frac{\Delta v_i}{v_{z0}}\sim 10\%$ is required for compression, which necessitates a maximum deflecting strength of the magnet of about $0.3\; T\cdot cm$ ($\Delta v= \frac{e\int B_r\cdot dl}{m}$).  In the ideal case of zero fringe field (FIG.\ref{FIG.2}(b)), the axial velocity of bunches remains constant in the drifting space. We solved for the appropriate time-varying deflecting strength $D(t)$ analytically (FIG.\ref{FIG.2}(c)). All of the compression component parameters above are achievable using existing technologies.
	
	In particular, the design of deflecting magnets differs from that of traditional bending magnets. The deflecting magnets are intended to generate local radial field within an uniform axial-oriented field $B_{z0}$.  The use of ferromagnetic yoke can introduce significant distortion, which would compromise the homogeneity of $B_{z0}$ and cause a lateral shift about $5 \; cm$ on bunches' spiral movement. Therefore, the yoke is not used in this case, and the fringe field of deflecting magnets is not absorbed(FIG.\ref{FIG.2}(b)). 
	
	Fringe field of the deflecting magnet affects the motion of electrons. Assuming that the i-th bunch is scheduled to pass through the magnet at time $t_i$. When it travels on the injection side ($t<t_i$), the fringe field could distract it from the flat field region of deflecting coil. And on the drifting space side ($t>t_i$), the fringe field could cause an axial track shift $\frac{\Delta z}{L}$ of about $3\% $ on bunch’s trajectory, which may result in extraction failure.
	
	 To address the issue caused by the fringe field, an extra steering coil with time-varying deflecting strength $O(t)$ is placed between the injection port and modulating magnet, as illustrated in the FIG.\ref{FIG.2}(d). The deflecting strength $D(t)$ of the modulating coil is also adjusted for compensation. Bunches' trajectories should satisfy two conditions for compression purpose:  arriving at the flat field region of modulating magnet, and reaching the demodulating magnet with zero axial shift.  The variables D(t) and O(t) should be carefully chosen to meet these criteria since the track shifts of the bunches $\Delta z_i$ are not linearly correlated with the deflecting strength. To solve for D(t) and O(t), we use an iterative algorithm.
	
\begin{figure}[htbp]
		\includegraphics[width=8.4cm]{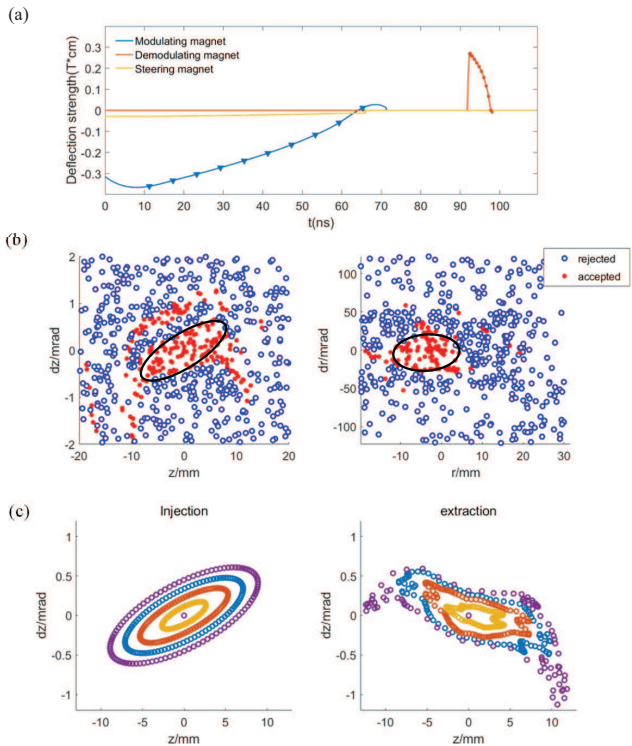}
		\caption{ (a) Calculated time-varying strength of steering and deflecting magnets. The triangle and circle symbols represent arrival time of bunches. (b) Phase-space acceptances for the first injected bunch. z is direction of static field $B_z$ and r is radial direction. (c) Transformation of first injected bunch’s phase-space distribution from injection into extraction port.}
		\label{FIG.2_2}
	\end{figure}

	Using the algorithm, we obtained a solution for the described set-up(FIG.\ref{FIG.2_2}(a)). Compression is still achievable considering the fringe field issue. In the simulated design, the time interval of compressed bunch train was narrowed to $T_b'=T_b-T_c=0.6ns$ after extraction. As a result, the current of the electron beam was multiplied by a factor of 10.
	
	Phase-space acceptances are calculated for this compression scheme, as shown in FIG.\ref{FIG.2_2}(b). The geometric transverse acceptance of the compression set-up is approximately $15 \; mm\cdot mrad$ in the z direction and $200 \; mm\cdot mrad$ in the r direction(laboratory coordinates). For the first injected bunch, which undergoes 17 turns in the drifting space, The compressing procedure causes a slight twist to its transverse phase-space distribution, as shown in FIG.\ref{FIG.2_2}(c), but with minimal emittance growth. To prevent bunch overlap or particle loss, an energy dispersion of $\pm 5 \%$ is allowed.  With the space-charge force taken into account, the electron bunch with an initial emittance of $1mm\cdot mrad$ and charge of up to 1nC is still accepted in the current design, indicating a compressed current of 1.85 A.

\begin{figure}[htbp]
		\includegraphics[width=9cm]{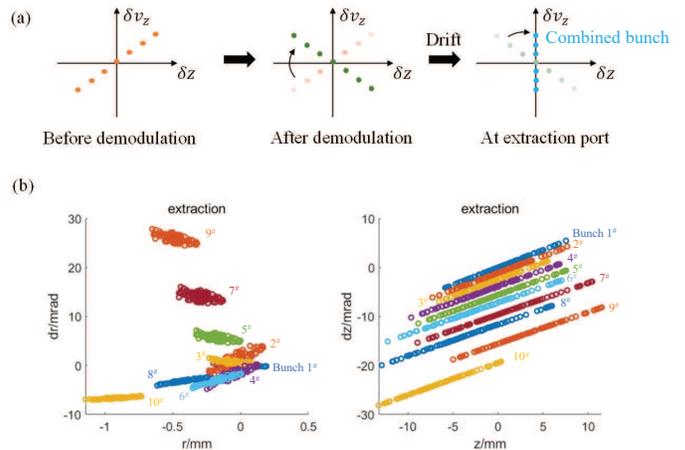}
		\caption{(a) Demodulation scheme in bunch combination system. (b) Phase-space distribution of combined bunch at extraction port.}
		\label{FIG.3}
	\end{figure}

	The compression ratio could be further increased by reducing difference between $T_b$ and $\Delta NT_c$. In the special case where $T_b=\Delta NT_c$, a combination of bunches can be made. In this combination case, all bunches reach the demodulating magnet at the same time but with different axial velocity. As shown in FIG.\ref{FIG.3}(a), a static quadrupole magnet is used as the demodulating magnet, which focuses the combined bunches in z-direction at the extraction port. The deflecting strength D(t) of the modulating magnet could be iteratively solved in a similar manner as mentioned above.
	
	The simulation of bunch combination has been performed with the same parameters illustrated in FIG.\ref{FIG.2}(a) and $T_b=T_c=5.4 \; ns$. The geometrical transverse acceptance is approximately  $1.2\; mm\cdot mrad$ in the z direction and $2\; mm\cdot mrad $ in the r direction. Energy acceptance is $\pm 5 \%$ relative to $6.5 \; MeV$. At the extraction port, a single electron bunch with a charge of 10 nC is obtained. The phase-space distribution of the combined bunch at the extraction port is shown in FIG.\ref{FIG.3}(c), where the injected bunches pile up one by one in phase space, as expected. Charge of combined bunch is mainly limited by space-charge expansion and the aperture of the extraction port.  It could be further improved by optimizing these components or introduce some focusing element in the drifting space.

	\section{Bunch confinement}
\begin{figure}[htbp]
		\includegraphics[width=9cm]{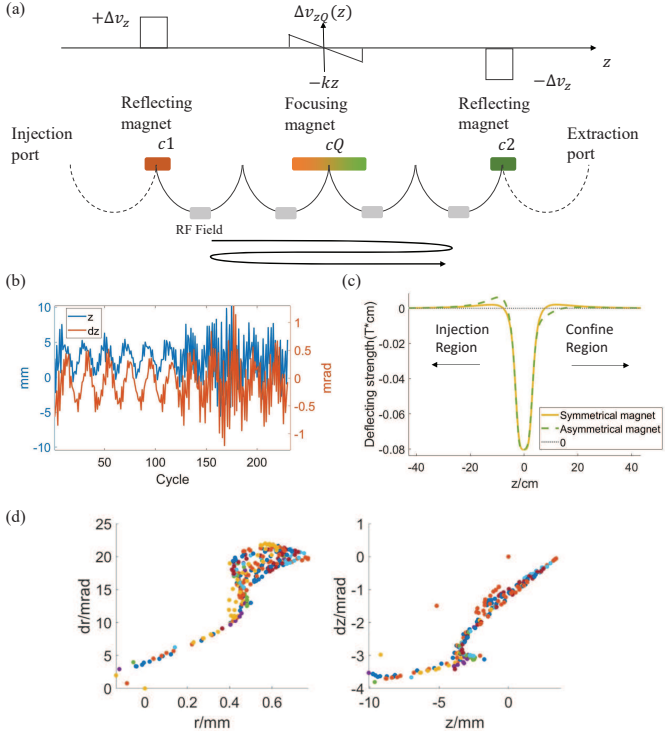}
		\caption{(a) Sketch of lattice design and bunch trajectory in confine scheme. (b)Position and momentum offset of confined bunch at each cycle. The offset value is measured at the focusing magnet. (c) Defelcting strengh of symmetrical and asymmetrical reflecting magnets. (e) Transvers phase-space distribution of the extracted bunches.}
		\label{FIG.4}
	\end{figure}

	The bunch confinement scheme, illustrated in FIG.\ref{FIG.4}(a), operates similarly to a magnetic mirror\cite{MagneticMirror}, where a pair of reflecting magnets can confine a bunch by reversing its axial velocity at the sides.  The left reflecting magnet can be switched for top-up injection. Ideally, this rapid switching will not impact the motions of the already-confined bunches. The charge of the electron beam in the confinement region can keep growing with incessant injection until the bunch interval is shorter than the magnet's switching duration.  Once the right reflecting magnet is turned off, all confined bunches flow out from the confinement region and are extracted with multiplied current. This scheme is analogous to a storage ring but is relatively compact. Furthermore, the system's huge dynamic aperture has potential for much larger storage current and significantly higher compressed power.	
	
	Similar to the lattice design of a storage ring, bunch focusing is neccessary. Quadrupole magnets are utilized to maintain the beam stability in the axial direction ($\vec{z}$). The uniform axial magnetic field provides weak focusing for the radial ($\vec{r}$) movements.  Additionally, several RF fields are implemented on bunches' confinement track for longitudinal ($\vec{s}$) focusing. The placement of focusing magnet and RF fields is illustrated in FIG.\ref{FIG.4}(a).

	Magnets' fringe field remains the obstacle in this confinement scheme, introducing instability to bunches' confined motion. If leaving out the injection and extraction issue of the confine system and keep all reflecting magnets on, a well-adjusted quadrupole magnet can maintain the shift of bunches' motion in the z direction within  $\pm 1 \; cm$ at reflect magnets position for no more than 250 confinement round trips (FIG.\ref{FIG.4}(b)). 
	
	Taking the injection and extraction issue account, the refelcting magnets are swithed on and off like kick magnets with a switch-off duration of  $1 \; ns$ for each injection event. This switching causes fluctuation of magnets' fringe field, which in turn deflects the existing confined bunches and leads to severe particle loss. To solve this concern, we have designed an asymmetrical magnet to reduce the flux of fringe field on the confinement region side(FIG.\ref{FIG.4}(c)).  Since the motion of bunches during injection or extraction is more controllable, fringe field enhancement ouside the confine region is acceptable, and its effect can be compensated for.  With the asymmetrical magnets, performance of confine system is improved and severe particle loss is avoided in simulation results.

	Installation in FIG.\ref{FIG.1}(b) is also employed in calculation of confinement scheme. Bunches take 8 turns for one round trip in the confine region. Time interval of the bunch train before and after compression are  $T_b = 83.7 \; ns$ and $T_b'=|T_b-8T_c|=2.7 \; ns$. The beam current is compressed by 31 times in this system.
	
	 The geometric transverse acceptance of the confinement scheme is approximately $1.2 \; mm\cdot mrad$ in the $\vec{z}$ direction and $55 \; mm\cdot mrad$ in the $\vec {r}$ direction. Energy acceptance of the confine structure reaches $\pm 5 \text{\textperthousand} $ with RF Field of 370 MHz, 1.25 kV. FIG. \ref{FIG.4}(e) shows transverse phase-space distribution of the extracted bunches, which had geometirc transverse emittance of $0.5 \; mm\cdot mrad$ and a gaussion profile at injection.  The bunches' envelope are twisted severely, but consistency among each bunch is good. 
	
	
	\section{Energy modualtion}
\begin{figure}[htbp]
		\includegraphics[width=9cm]{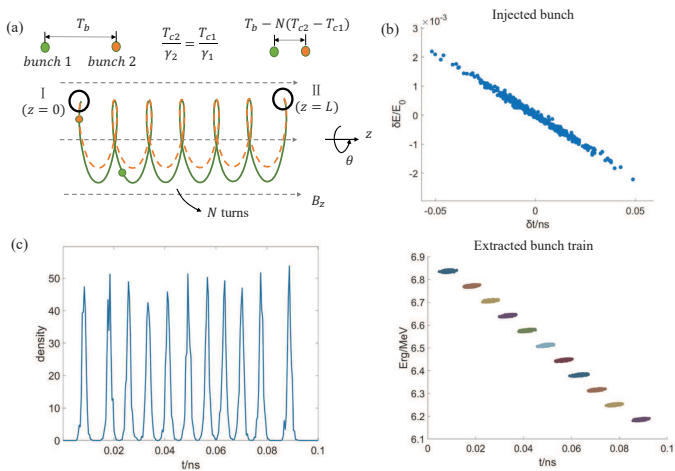}
		\caption{(a) Sketch of the compression scheme. In the uniform magnetic field, Bunch 2 spirals with higher energy ($\gamma _2 > \gamma _1$) and longer period ($T_{c2} > T_{c1}$). Bunches' interval turns into $T_b' = T_b - N(T_{c2}-T_{c1})$ at position \uppercase\expandafter{\romannumeral2}. (b)Longitudinal phase-space distribution of each injected bunch and extracted bunch train. (c)Normalized electron density of the compressed beam at the extraction port.}
		\label{FIG.5}
	\end{figure}

	As the turning period of an electron in a uniform magnetic field is proportional to its energy ($T_c \propto E$), the interval between bunches can be adjusted by modulating their energy before injection. Sketch of this compression scheme is presented in FIG. \ref{FIG.5}(a).			
	
	With pre-modulated bunch train, electromagnet is needless inside the cylindrical cavity. This eliminates issues associated with fringe field or magnet power supply. 
	
	Given the aforementioned  $0.1 \; T$ magnetic field and $6.5 \; MeV$ bunch train, a relative energy dispersion of $ 1\% $ would cause a time delta of $54 \; fs$ for each sipral turn. It means a finer manipulation on time scale, and has potential to compress electron beam to a W-Band repetition rate .
	
	We conducted preliminary calculations of 10 bunches with $T_b=0.52 \; ns$. Each bunch's energy was sequentially decreased by $1\%$ relative to $E_0 =6.5 \; MeV$.  After 10 laps of sapiral motion inside the cavity, we successfully formed a bunch train with a frequency of 100 GHz at the extraction port.
	
	Due to the extremely short interval of extracted bunches,  technical difficulties arise, including bunch longitudinal size growth caused by energy spread within each bunch, and system's nonlinearity of time delay with beam energy. The former problem can cause an overlap of adjacent bunches, while the latter results in heterogeneity of the extracted bunch interval. An RF field applied at the center of the spiral trajectories could adress both the difficuties to some extent. In addition to the bunching effect, it also supplies negative feedback regulation for the bunch interval.  In simulation, RF Field of 7.6 GHz, $70 \; kV$ is employed.
	
	Longitudinal phase-space distribution of the injected bunch and the extracted beam is shown in FIG. \ref{FIG.5}(b). The average pulse width of each current peak is a mere 6 ps. And the waveform has a form factor of approximately 0.5 (FIG. \ref{FIG.5}(c)). 
	
	\section{conclusion}
	
	In this letter, we demonstrate a macro electron beam compression concept achieved by modulating bunches' spiral motion in a uniform magnetic field. With proposed set-up. Simulation of 3 different compression schemes are performed. 
	
	In spiral pitch modulation scheme, compressed beam with 10 times multiplied current of 1.85A is obtained. And in bunch confinement scheme, compression of a beam with initial duration of 838 ns is implemented in the several-meters-long set-up. 
	
	It composing a economy solution for manipulating electron beam on the timescale of nanoseconds to microseconds. 

	\bibliography{ref}
	
\end{document}